\documentclass[runningheads]{llncs}
\usepackage{graphicx}
\usepackage{float}
\usepackage{xcolor}

\usepackage{svg}
\usepackage{subfig}
\usepackage{graphicx}

\begin{document}
%


\title{Hope Amid of a Pandemic: \\
Is Psychological Distress Alleviating in South America while Coronavirus is still on Surge?}

\titlerunning{Pyschological Distress, South America, Coronavirus}
%
\author{Josimar E. Chire-Saire\inst{1}\orcidID{0000-0003-2942-0853} \and Khalid Mahmood \inst{2}
}
\authorrunning{Chire Saire, J. and Mahmood, K.}
%

\institute{Institute of Mathematics and Computer Science (ICMC),\\
University of São Paulo (USP),\\ São Carlos, São Paulo, Brazil\\
\email{jecs89@usp.br}
\and
Department of Information Technology,\\
Uppsala University,\\
Uppsala, Sweden\\
\email{khalid.mahmood@it.uu.se}}

\maketitle              
\begin{abstract}
    As of July 31, 2020, the COVID-19 pandemic has over 17 million reported cases, causing more than 667,000 deaths. Countries irrespective of economic status have succumbed to this pandemic. Many aspects of the lives, including health, economy, freedom of movement have been negatively affected by the coronavirus outbreak. Numerous strategies have been taken in order to prevent the outbreak. Some countries took severe resections in the form of full-scale lockdown, while others took a moderate approach of dealing with the pandemics, for example,  mass testing, prohibiting large-scale public gatherings, restricting international travels. South America adopted primarily the lockdown strategies due to inadequate economy and health care support.  Since the social interactions between the people are primarily affected by the lockdown,  psychological distress, e.g. anxiety, stress, fear are supposedly affecting the South American population in a severe way. This paper aims to explore the impact of lockdown over the psychological aspect of the people of all the Spanish speaking South American capitals. We have utilized infodemiology approach by employing large-scale Twitter data-set over 33 million feeds in order to understand people’s interaction over the months of this on-going coronavirus pandemic. Our result is surprising: at the beginning of the pandemic, people demonstrated strong emotions (i.e. anxiety, worry, fear)  which declined over time even though the actual pandemic is worsening by having more positive cases, and inflicting more deaths. This leads us to speculate that the South American population is adapting to this pandemic thus improving the overall psychological distress.  
 

\end{abstract}

\section{Introduction}

The COVID-19 pandemic is one the of modern day’s calamity which currently affects the population of almost every country in the world. The disease caused by the \textit{Severe Acute Respiratory Syndrome Coronavirus 2 (SARS‑CoV‑2)} a.k.a \textit{coronavirus}, which is first identified in the Chinese city of Wuhan. Just over a month, the pandemic is recognized as \textit{Public Health Emergency of International Concern} by the World Health Organization (WHO) \cite{world2020statement}. As of July 31, this virus infected over 17 million people  causing over 672,000 deaths. 

The coronavirus demonstrated a fast growth of infections within and outside of China and soon spread nearby Asian countries, Europe and North America. This virus overwhelmed the sophisticated public health infrastructures of the developed countries, e.g. Italy, Spain, the United Kingdom, and the United States, consequently took many lives. 

Numerous public safety measures have been taken in order to prevent the fast spreading of the virus. Some countries utilized severe forms of measurement such as full-scale lockdown strategy while others utilized moderate form of action including mass testing, drive through testing, social distancing, limiting mass gathering, travel restriction, etc. These strategies by no means have stopped the on-going pandemic.

South America, first observed it's case on 26 February 2020 in Brazil. By June 26, it has more than 2 million confirmed cases consisting of 81,000 deaths, eventually became on the main epicenter of coronavirus crisis in the world. 

Being a continent consisting of mostly developing countries, South America faces a unique challenge in dealing with the pandemic. Deprived health care system and limited testing capability\footnote{Uruguay are one of the few countries in South  American to limit the outbreak due to extensive testing.}  are among the main reasons for not dealing with COVID-19 crisis successfully. The primary measurement of most of the South American capitals to deal with the coronavirus crisis is to limit the restriction of human movement, e.g. partial to full-scale lockdown, prohibiting large-scale public gathering, suspending schools. Since the lockdown is not sustainable and in fact, greatly impacting the livelihood of the people of developing countries, eventually it’s needed to be lifted over in order to bolster the economy. 

Apart from the health and economic crisis due to the COVID-19 outbreak, limiting the \textit{freedom of movement} significantly impacts our psychological well-being. Amid of lockdown, people are more vulnerable towards the apprehension of the unforeseeable future. Therefore, understanding the psychological hardship of the South American population during this pandemic will be a great means to facilitate the judicious public health decisions during this ongoing pandemic which is likely to be prevalent for some time period.

\textit{Infodemiology}\cite{Gunther2011} is a new research field, with the objective of monitoring public health\cite{Maged2010} in order to support public policies based on electronic sources, i.e. Internet. In infodemiology, the data that is used in research is primarily open, and textual having no structure and comes from different internet services, e.g. blogs, social networks, and websites. 

One of the key aspects of making rapid and effective public health actions is to concurrently observe the situation that demanded a fast response. Infoveillance approach could perform real-time monitoring of the situation by utilizing so-called \textit{social sensors}\cite{wang2018},  where individual or communities of people share their present status through different social media platforms (e.g. Facebook, Twitter). The data from social media could effectively be utilized in real-time in order to make adaptive public health measures to alleviate crisis demonstrating unpredictability in nature such as a pandemic or a disease outbreak.  The infoveillance approach, tailored towards surveillance proposals has previously been applied to monitor H1N1 pandemic with data source from Twitter \cite{chew2010pandemics}, a Dengue outbreak in Brazil using Social Sensors and Natural Language Processing\cite{chire2019}, COVID-19 symptoms in Bogota, Colombia using Text Mining \cite{saire2020people}.

In this paper, we are trying to understand \textit{how COVID-19 is impacting the psychological health of the population in Spanish speaking South American capitals}. We are utilizing the infodemiologyic approach to understand our research question by analyzing social media data sets. In summery, the \textbf{contributions of the paper} are as follows:
\begin{itemize}
    \item We explore how the population living in the Spanish speaking South American capitals facing the psychological distress during COVID‑19 pandemic. We demonstrated that the Infoveillance approach of analyzing social media data, i.e. Twitter, can be effectively utilized to understand people's mental status during the pandemic. Surprisingly, this has not been considered before. 
    \item We have collected a large number of  Twitter feeds of all the Spanish speaking South American capitals which are cumulatively around 33 million of data points and applied proper data mining methods in order to effectively understand our research question (discussed in section \ref{Chap:analysis}).
    \item Our results described in Section \ref{Chap:twitter} is counter-intuitive and surprisingly demonstrated that people's interest in the pandemic is decreasing, inferring that the psychological distress is elevating over the months while the actual pandemic is worsening (discussed in section \ref{sec:discussion}). \textit{This is the most important finding of our work.}
    \item In section \ref{Chap:trend}, we have verified our findings with another social media platform, i.e. \textit{Google Trend}, and demonstrated that similar pattern can also be observed in \textit{Google Search}.
    \item We discuss our overall finding and speculate that this can be a potential starting point to perform comprehensive research to explore the infoveillance approach for understanding the psychological impact due to COVID-19 pandemic.
\end{itemize}

This paper is structured as follows: section \ref{Chap:analysis} explains the Data Process Analysis for our experiments, section \ref{Chap:Results} presents results and analysis. Section \ref{sec:discussion} provides the discussion and Section \ref{Chap:conclusion} states the conclusions.


\section{Data Process Analysis} \label{Chap:analysis}

\begin{figure}[hbpt]
\centerline{
\includegraphics [width=0.9\textwidth]{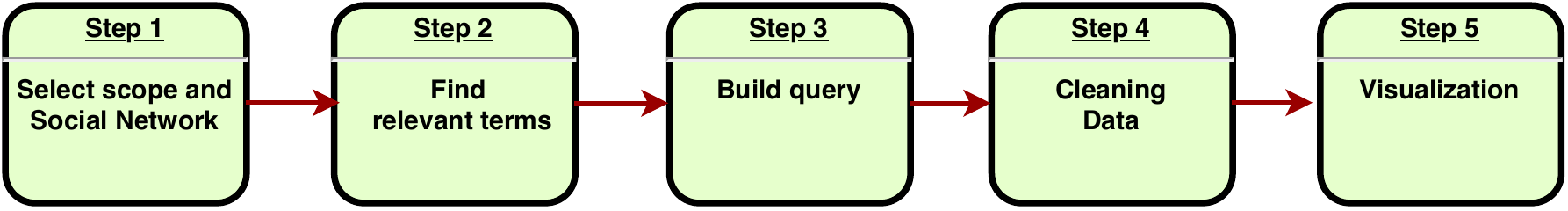}
}
\caption{Data Process Analysis}
\label{fig:data_workflow}
\end{figure}

The present analysis is inspired by the \textit{Cross Industry Standard Process for Data Mining(CRISP-DM)} \cite{crisp2000} steps, which consists of the frequent phases of Data Mining tasks. The workflow for data analysis is presented in Fig. \ref{fig:data_workflow} consists of five steps. 

The following subsections explain every step from data collection to visualization.

\subsection{Selecting the Scope and Social Network}

South America, being one of the most impacted continents of the pandemic, consists of 10 countries where official languages are either Spanish or Portuguese. Among those, Brazil is the only county where the official language is Portuguese. The rest of 9 countries in South America where Spanish is the official language are Argentina, Bolivia, Chile, Colombia, Ecuador, Paraguay, Perú, Uruguay, Venezuela.  Table \ref{tab:sainfo} presents the corresponding data related to Spanish spoken countries.
\begin{table}[H]
\centering
\caption{Cartographic and Demographic Information of the Spanish Speaking South American Countries}
\begin{tabular}{|r|c|c|c|c|}
\hline
\textbf{Country} & \textbf{Capital} & \textbf{Area($km^2$)} & \textbf{Population} & \textbf{People/$km^2$}                     \\ \hline
Argentina     & Buenos Aires     & 2,792,600           & 44,938,712          &  16.092 \\ \hline
Bolivia       & La Paz           & 1,098,581           & 11,383,094          &  10.361 \\ \hline
Chile         & Santiago         & 756,102             & 19,107,216          &  25.270 \\ \hline
Colombia      & Bogotá           & 1,141,748           & 50,372,424          &  44.118 \\ \hline
Ecuador       & Quito            & 283,561             & 17,300,000          &  61.009 \\ \hline
Paraguay      & Asunción         & 406,752             & 7,152,703           &  17.584 \\ \hline
Perú          & Lima             & 1,285,216           & 32,950,920          &  25.638 \\ \hline
Uruguay       & Montevideo       & 176,215             & 3,529,014           &  20.026 \\ \hline
Venezuela     & Caracas          & 916,445             & 28,067,000          &  30.625 \\ \hline
\end{tabular}
\label{tab:sainfo}
\end{table}



The analysis is conducted over the capitals of each country because there is more concentration of population having affordable access to the internet compare to other cities. In this context, the urban population act as \textit{social sensors} can potentially be utilized to inspect an event using interaction in the social media platform through posts/publications. Twitter is the choice of social network used in this work since the population of all the Spanish speaking countries uses it. Users on Twitter can send messages up to 240 characters long. It also facilities and access to data through it’s \textit{Application Programming Interface(API)}, which data scientists prefer for the ease of collecting the data.



\subsection{Find the Relevant Terms to Search}

Initially, we are interested in finding the twitter feeds related of coronavirus pandemic. In order to perform this, we have chosen terms that are related to coronavirus such as: \textit{coronavirus}, \textit{covid19}, etc. Since the users do not follow a specific pattern to a keyword when writing a post, where \textit{@coronavirus}, \textit{\#covid-19}, \textit{\#covid\_19} all similar to keyword COVID-19, we have considered variations of having special characters (i.e. \textit{@}, \textit{\#}, \textit{\textendash}, \textit{\_}) in our search terms. Similar strategy has also been utilized in the work of \cite{chire_south}.

\subsection{Build the Query to Collect Twitter Data}

The query to extract tweets using API uses the next parameters:

\begin{itemize}
    \item \texttt{date}: 01-04-2020 to 30-06-2020
    \item \texttt{terms}: the chosen words mentioned in previous subsection
    \item \texttt{geolocalization}: the longitude and latitude of 9 Spanish speaking capitals
    \item \texttt{language}: Spanish
    \item \texttt{radius}: 50 $km$
\end{itemize}

\subsection{Preprocessing of the Data}
We have performed  the following pre-processing steps in order to filter out unwanted information that could impact our analysis.

\begin{itemize}
    \item Cleaning \textit{urls} using regular expressions
    \item Eliminating \textit{non-alphabet} characters
    \item Converting all text to \textit{lowercase} only letters  
    \item Deleting \textit{stopwords}, e.g. articles, conjunctions
\end{itemize}

\subsection{Visualization}
When we are considering visualization, our paper depicts data from two different sources:
\begin{itemize}
    \item \textit{Twitter data}: filtering with 3 specific keywords related to people’s psychological distress  which are \textit{anxiety}, \textit{fear} and \textit{stress}.
    \item \textit{Google Trend's data}: for displaying what is trendy for a particular keyword in google search for our chosen countries. 
\end{itemize}

\section{Results} \label{Chap:Results}

\subsection{Dataset Description}

The collected data has three main attributes: \textit{date}, \textit{text}, \textit{user\_name}. The Table \ref{tab:summary} provided overview of the collected 33 million Twitter feeds from 9 different capital cities.

\begin{table}[]
\caption{Dataset Description of 9 South American Capital Cities}
\label{tab:summary}
\centering
\begin{tabular}{|r|c|c|c|}
\hline
\textbf{Country}                & \textbf{Number of tweets} & \textbf{Unique users} & \textbf{Tweets/user} \\ \hline
Argentina                       & 5,973,746                 & 737,765                & 8.097                \\ \hline
Bolivia                         & 260,405                   & 21,364                 & 12.189               \\ \hline
Chile                           & 5,616,438                 & 303,700                & 18.493               \\ \hline
Colombia                        & 3,192,229                 & 310,816                & 10.270               \\ \hline
Ecuador                         & 1,887,503                 & 100,911                & 18.704               \\ \hline
Perú                            & 4,312,931                 & 284,621                & 15.153               \\ \hline
Paraguay                        & 2,981,195                 & 154,301                & 19.321               \\ \hline
Uruguay                         & 1,402,344                 & 188,328                & 7.446                \\ \hline
Venezuela                       & 7,905,287                 & 297,685                & 26.556               \\ \hline
\textbf{Total} & 33,532,078                & 2,399,491              & 15.137               \\ \hline
\end{tabular}
\end{table}

From the table, it is evident that the two countries has most Twitter feeds, are Venezuela and Chile, whereas Bolivia and Uruguay consists the least. On the other hand, Paraguay is having the average feeds among South American countries.

\subsection{Evolution of Interest on Twitter related to Psychological Distress} \label{Chap:twitter}

\begin{figure}[H]
\centerline{
\subfloat[Argentina]{%
\includegraphics[width=0.7\textwidth]{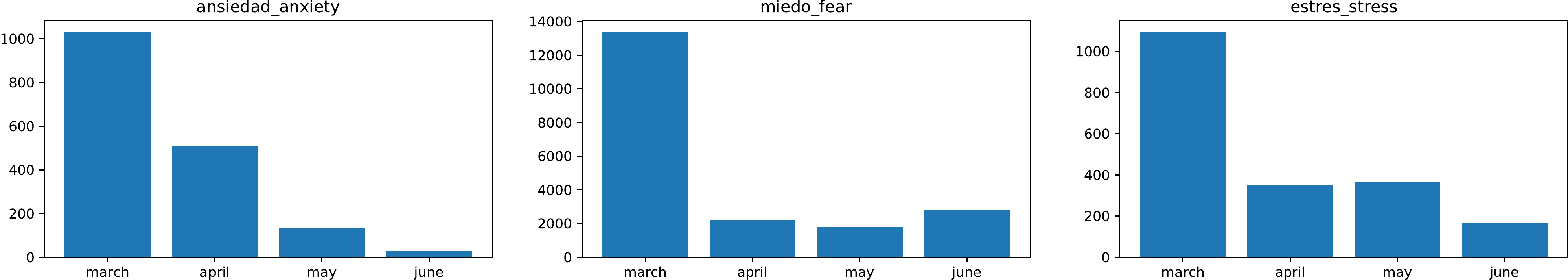}
}
}

\centerline{
\subfloat[Bolivia]{%
\includegraphics[width=0.7\textwidth]{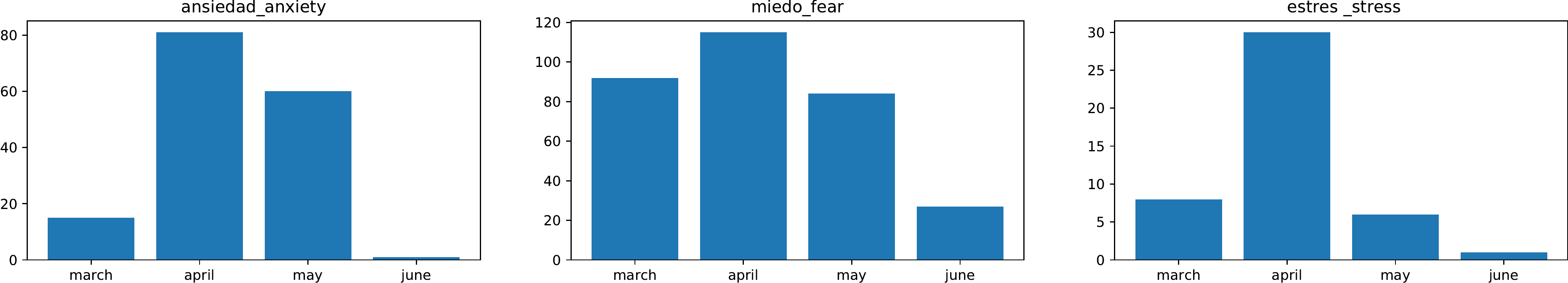}
}
}

\centerline{
\subfloat[Chile]{%
\includegraphics[width=0.7\textwidth]{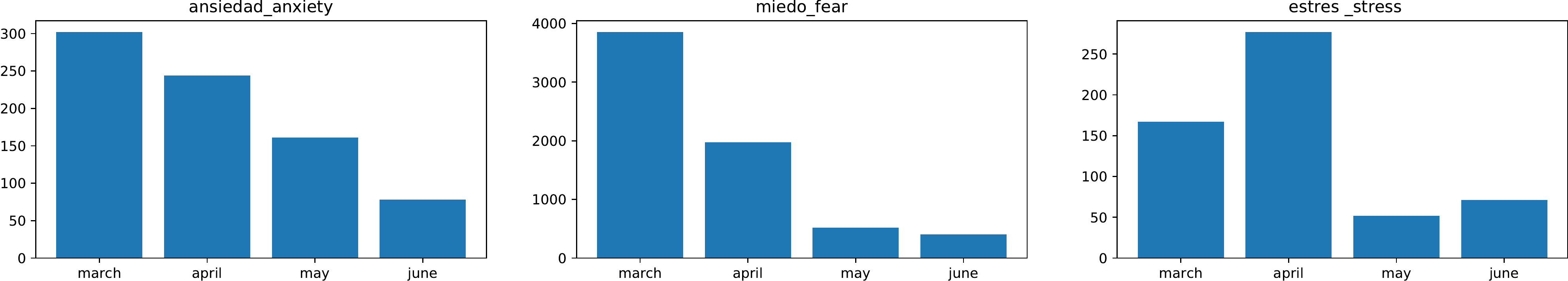}
}
}
\caption{Evolution of terms related to phycological distress for the capitals of Argentina, Bolivia, and Chile}
\label{fig:countries1}
\end{figure}

\begin{figure}[H]
\centerline{
\subfloat[Colombia]{%
\includegraphics[width=0.7\textwidth]{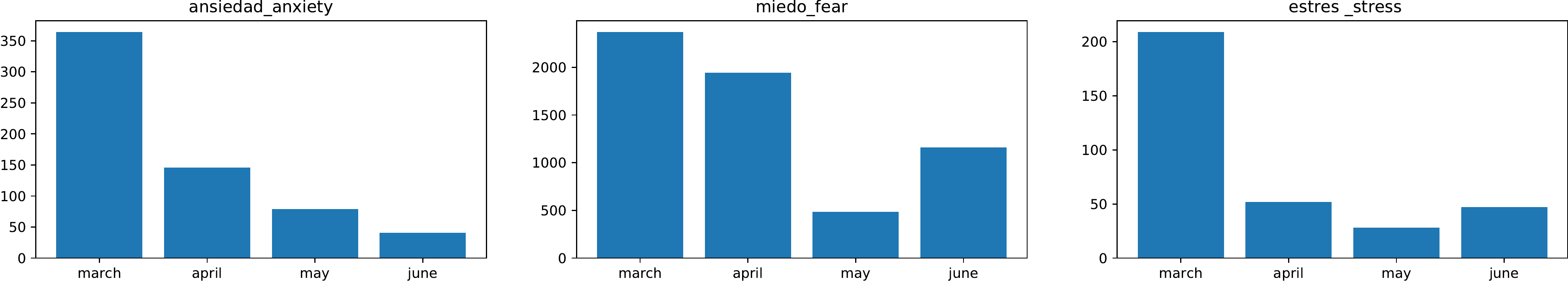}}
}
\centerline{
\subfloat[Ecuador]{%
\includegraphics[width=0.7\textwidth]{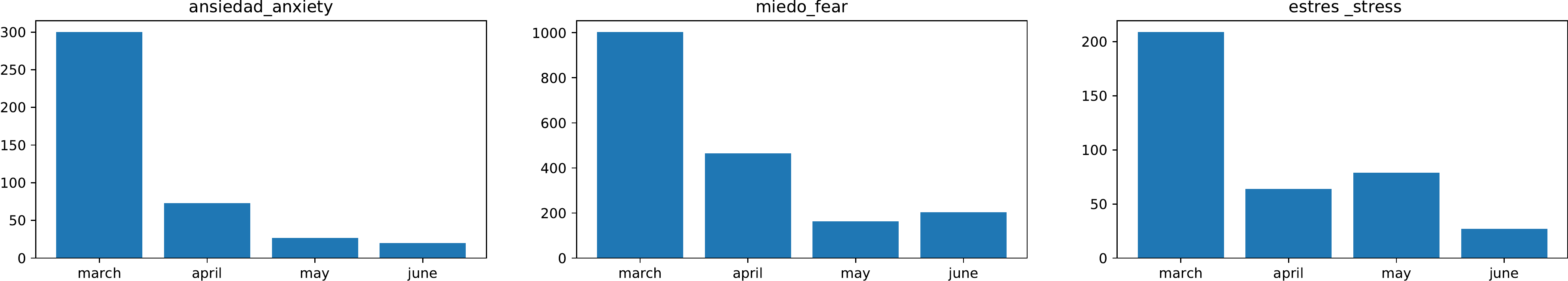}
}
}
\centerline{
\subfloat[Peru]{%
\includegraphics[width=0.7\textwidth]{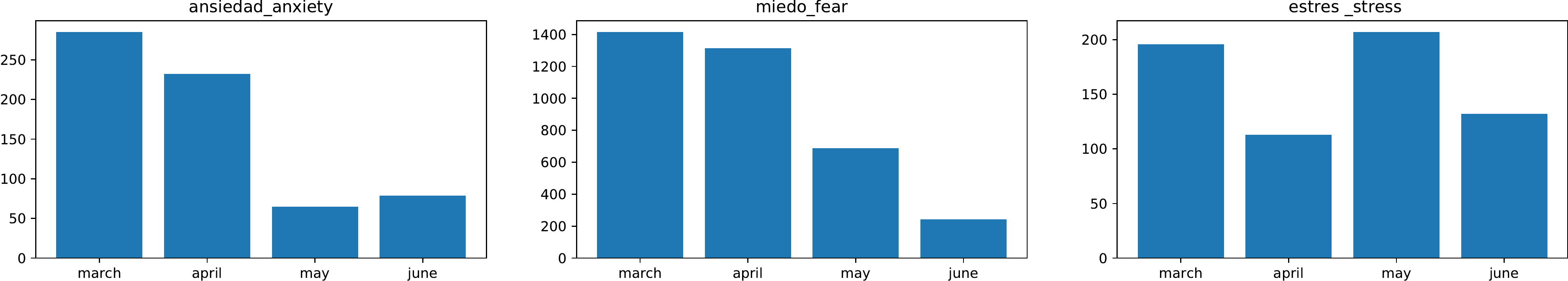}
}
}
\caption{Evolution of terms related to phycological distress for the capitals of Colombia, Ecuador, and Peru}
\label{fig:countries2}
\end{figure}

\begin{figure}[H]
\centerline{
\subfloat[Paraguay]{%
\includegraphics[width=0.7\textwidth]{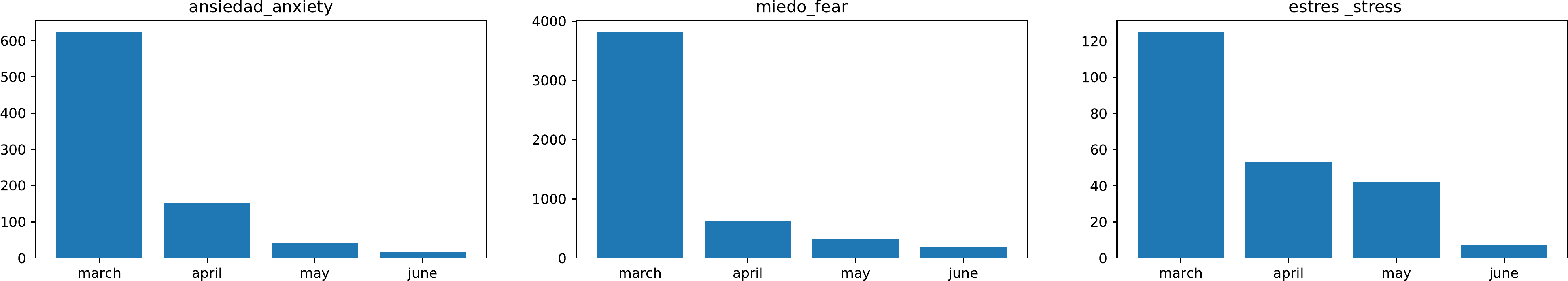}}
}
\centerline{
\subfloat[Uruguay]{%
\includegraphics[width=0.7\textwidth]{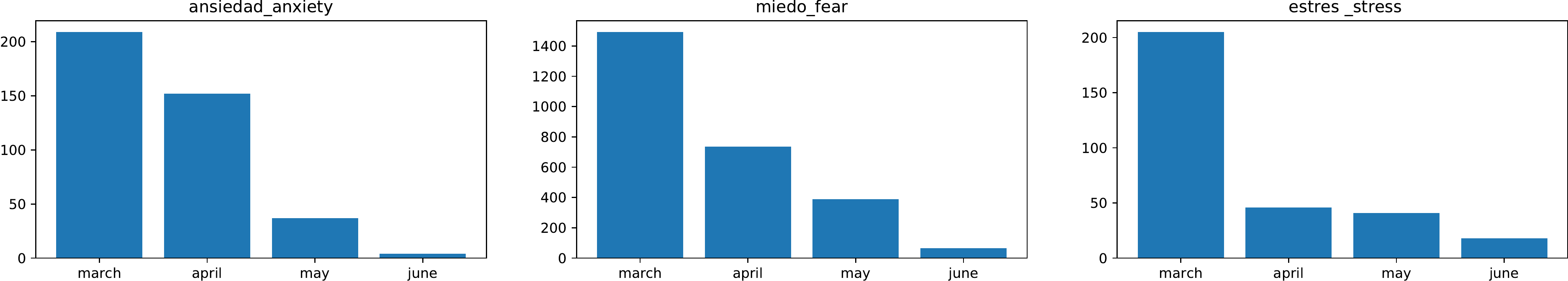}}
}
\centerline{
\subfloat[Venezuela]{%
\includegraphics[width=0.7\textwidth]{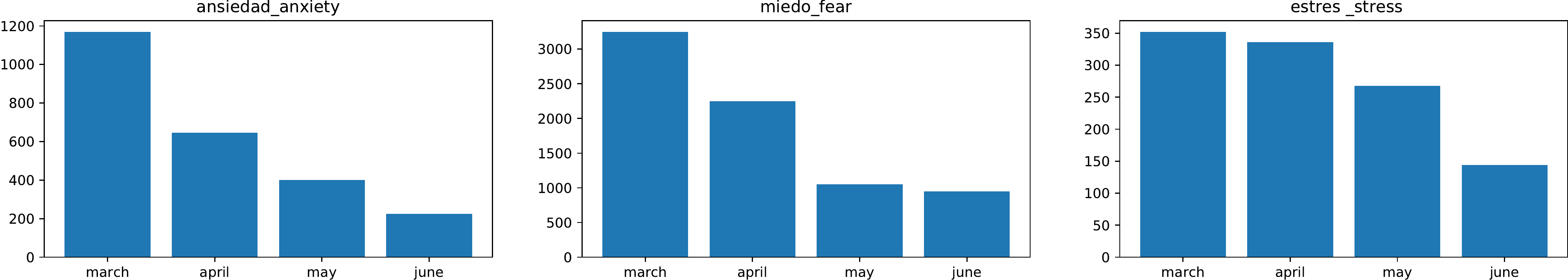}}
}
\caption{Evolution of terms related to phycological distress for the capitals of Paraguay, Uruguay, and Venezuela}
\label{fig:countries3}
\end{figure}

In this section, we are exploring how the psychological distress that related to our three terms \textit{anxity}, \textit{fear}, and \textit{stress} evolve over four months of time from March to June during this pandemic for the 9 capital cities. We have divided 9 capitals among 3 different figures having 3 capitals each which are presented in Fig. \ref{fig:countries1},\ref{fig:countries2},\ref{fig:countries3}. In these figures, we present the months from March until June in the x-axis, while in the y-axis, we have reported the corresponding frequency that is appeared for the corresponding psychological terms for that particular month.

Fig. \ref{fig:countries1} presents the scenario for Argentina, Bolivia and Chile. Argentina and Chile present a decreasing behavior from March to June for both terms: \textit{anxiety} and \textit{fear}. Bolivia reaches its peak in April followed by a downward pattern for all three terms. For term, \textit{stress},  Argentina has a clear decrease from March and on-ward while Chile demonstrates similar behavior to Bolivia. 

Fig. \ref{fig:countries2}, presents the results for Colombia, Ecuador, and Peru. The three countries present a decreasing number of posts during the four months of the pandemic for three terms. An exception is evident in Peru, a peak in May related to \textit{stress} term can be observed.

Fig. \ref{fig:countries3} introduces the outcome for Paraguay, Uruguay and Venezuela. All three countries demonstrated a decreasing trend in frequency of posts from March to June for the corresponding three terms: \textit{stress}, \textit{anxiety} and \textit{fear}.

One clear conclusion can be made that all South American countries have a decreasing trend for these three terms related to psychological distress.

\subsection{Validating the Analysis using Google Trend} \label{Chap:trend}

In order to validate our finding, we have considered the \textit{Google Trends} to have a perspective about how South American people’s interest over coronavirus generally appeared in the \textit{Google Search}.

\begin{figure}[H]
\centerline{\includegraphics [width=1.0\textwidth]{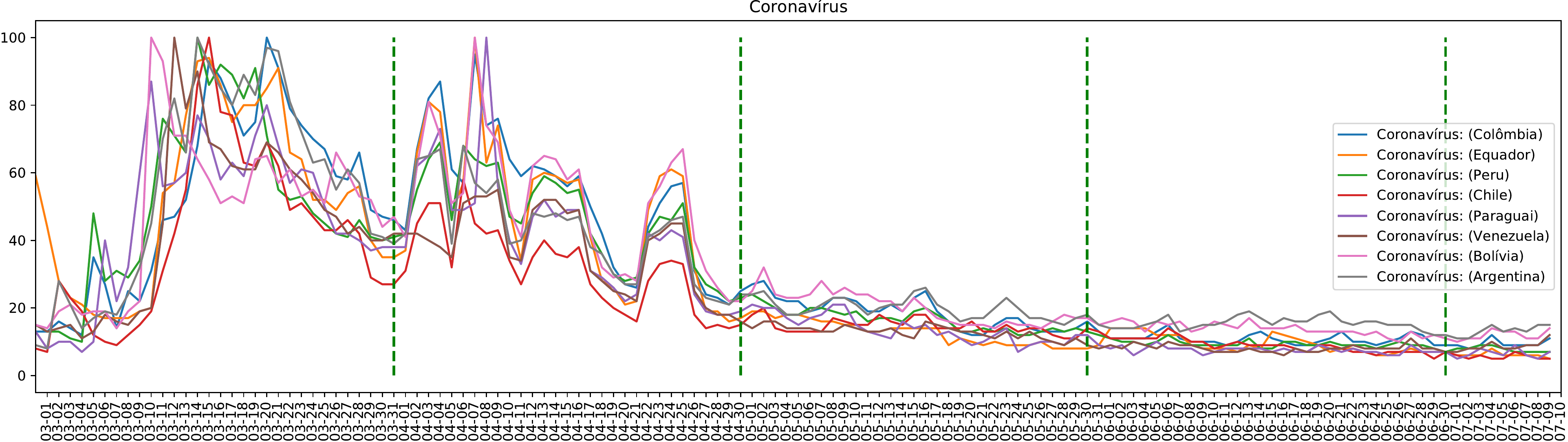}}
\caption{Google Trends with search term coronavirus (vertical dashed lines separates the months)}
\label{fig:trends}
\end{figure}

In Fig. \ref{fig:trends}, we have displayed the frequency of \textit{coronavirus} search term appeared in \textit{Google Search} during March until July, relevant to our 9 countries.

Surprisingly, all the countries present a decreasing interest of term \textit{coronavirus} which is similar to our previous analysis using Twitter data.

In fact, it is quite surprising that the interest of the South American population are similar when it comes to the interest in coronavirus and almost all the country follows a similar trend.

\section{Discussion} \label{sec:discussion}

\begin{figure}[H]
\subfloat[Daily new cases]{%
\includegraphics[width=0.45\textwidth]{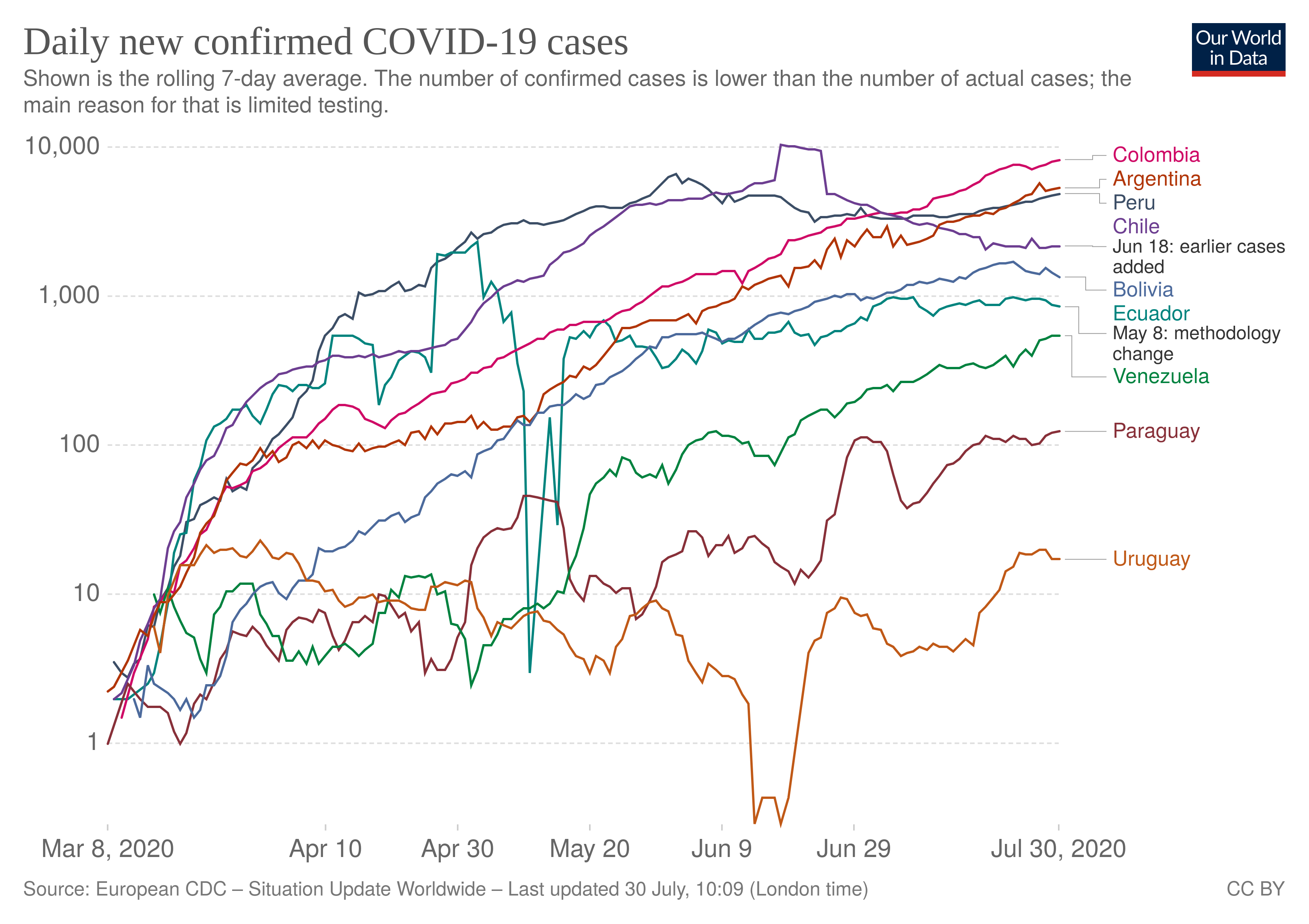}
}
\subfloat[Daily new death]{%
\includegraphics[width=0.45\textwidth]{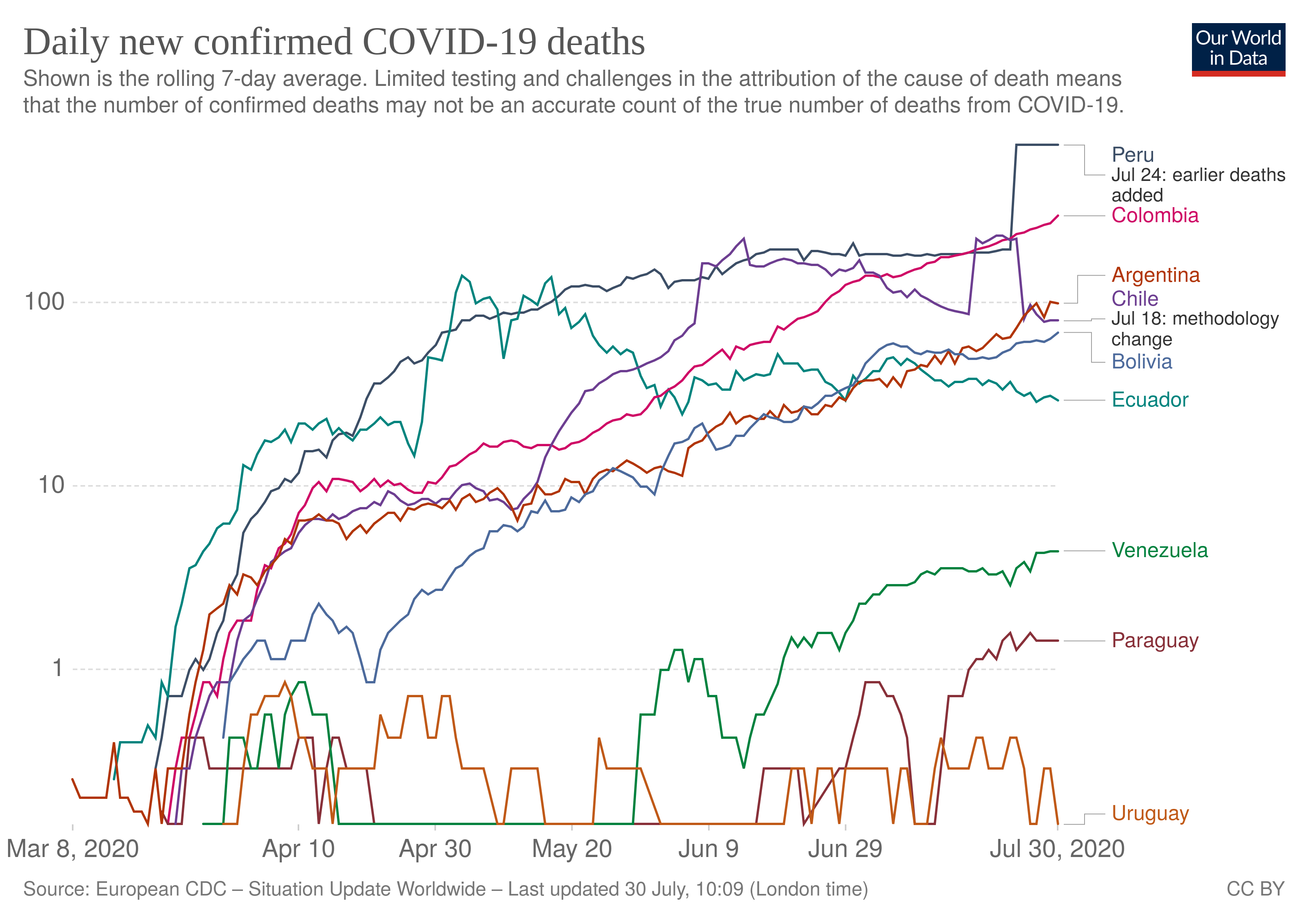}
}
\caption{Daily new cases and Deaths due to COVID-19 in 9 South American Countries\protect\footnotemark} 
\label{fig:countries4}
\end{figure}

\textit{Are people in South America have been accustomed to the pandemic?} In order to understand such phenomena, in Fig. \ref{fig:countries4}, we have depicted the daily \textit{new cases} and \textit{deaths} corresponds to the COVID-19 pandemic for the month of March until July 2020. 

\footnotetext{The graphic is generated using a platform from www.ourworldindata.org}

From Fig. \ref{fig:countries4}, it is clearly evident that in most South American countries, both the total number of new cases and death are low at the beginning which increases over time. Interestingly, at the beginning of March/April, most countries impose a stricter restriction on people’s \textit{freedom of movement}, such as forced lockdown, however, people’s psychological distress was high  (demonstrated in two previously Subsections \ref {Chap:twitter} and \ref{Chap:trend}) even having relatively low number of active cases and deaths. As the coronavirus pandemic exacerbates by having more positive cases and deaths, people’s apprehension about the coronavirus is surprisingly alleviated. This might properly explain why enforced confinement enables psychological distress upon the population. 

\section{Conclusions} \label{Chap:conclusion}
In this paper, we have employed large scale social-media data (i.e. Twitter) in order to understand the psychological distress due to COVID-19 pandemic of the Spanish speaking South American population.  We have found that even though coronavirus pandemic is aggravating with having more active cases and deaths, people's interaction in social media related to anxiety, worry, and fear of coronavirus is decreasing. We are speculating that people are becoming more accustomed to the pandemic compare to the beginning. This leads us to an optimistic conclusion that the people mental distress is alleviating reflecting the innate nature of humanity for overcoming tough-time.

\bibliographystyle{apalike}
\bibliography{biblio.bib}





\end{document}